\documentclass[journal]{IEEEtran}
\usepackage[utf8]{inputenc}
\usepackage{blindtext}
\let\labelindent\relax
\usepackage{enumitem}
\usepackage{csquotes}
\usepackage{biblatex} 
\usepackage{graphicx}
\usepackage{float}
\usepackage{amsmath}
\usepackage{hyperref}
\graphicspath{ {images/} }
\usepackage{csvsimple}
\usepackage{colortbl}
\usepackage[table,xcdraw]{xcolor}
\usepackage{multirow}

\usepackage[english]{babel}
\usepackage{amssymb}
\usepackage{xcolor}
\usepackage{soul}
\usepackage{algorithm}
\usepackage{algpseudocode}

\bibliography{citation.bib}
\usepackage{caption}

\ifCLASSINFOpdf

\else
\fi
\begin{document} 
%
\title{\LARGE 
Bidirectional Brain Image Translation using Transfer Learning \\from Generic Pre-trained Models}
\author{Fatima Haimour$^{1}$, Rizik Al-Sayyed$^{2}$, Waleed Mahafza$^{3}$ and Omar S. Al-Kadi$^{2}$\\
\textit{$^{1}$Faculty of Information Technology, Zarqa University, Zarqa 13110, Jordan \\$^{2}$King Abdullah II School for Information Technology,\\ University of Jordan, Amman 11942, Jordan\\$^{3}$Department of Diagnostic Radiology, Jordan University Hospital, Amman 11942, Jordan}}
\maketitle

\begin{abstract}
 Brain imaging plays a crucial role in the diagnosis and treatment of various neurological disorders, providing valuable insights into the structure and function of the brain. Techniques such as magnetic resonance imaging (MRI) and computed tomography (CT) enable non-invasive visualization of the brain, aiding in the understanding of brain anatomy, abnormalities, and functional connectivity. However, cost and radiation dose may limit the acquisition of specific image modalities, so medical image synthesis can be used to generate required medical images without actual addition. CycleGAN and other GANs are valuable tools for generating synthetic images across various fields. In the medical domain, where obtaining labeled medical images is labor-intensive and expensive, addressing data scarcity is a major challenge. Recent studies propose using transfer learning to overcome this issue. This involves adapting pre-trained CycleGAN models, initially trained on non-medical data, to generate realistic medical images. In this work, transfer learning was applied to the task of MR-CT image translation and vice versa using 18 pre-trained non-medical models, and the models were fine-tuned to have the best result. The models' performance was evaluated using four widely used image quality metrics: Peak-signal-to-noise-ratio, Structural Similarity Index, Universal Quality Index, and Visual Information Fidelity. Quantitative evaluation and qualitative perceptual analysis by radiologists demonstrate the potential of transfer learning in medical imaging and the effectiveness of the generic pre-trained model. The results provide compelling evidence of the model's exceptional performance, which can be attributed to the high quality and similarity of the training images to actual human brain images. These results underscore the significance of carefully selecting appropriate and representative training images to optimize performance in brain image analysis tasks.
\end{abstract}


\IEEEpeerreviewmaketitle
\section{Introduction}
Brain images play a critical role in the medical image scarcity and assessment of brain anatomy, functions, and abnormalities~\cite{1}. Techniques such as magnetic resonance imaging (MRI) and computed tomography (CT) provide valuable diagnostic information for various neurological conditions, aiding in treatment planning and monitoring patient's progress~\cite{2}. Each modality has its own unique strengths and limitations~\cite{3}. The choice of modality depends on the clinical problem, patient's condition, practitioner's medical expertise, cost, radiation dose, patient age, and the limited availability of images~\cite{4,5}.

To address these concerns, medical image synthesis can assist with the generation of necessary medical images without the need for physical scanning ~\cite{6,18}. This technique can be especially helpful when imaging is not readily available or when there are concerns about the risks associated with radiation exposure, such as in the case of pregnant women or young children. CT imaging, for example, emits radiation, while MR imaging does not ~\cite{7}. However, in some cases, both modalities may be required to get a more complete picture of the patient's condition. Additionally, CT-MR image synthesis is used as a preliminary step in medical image segmentation ~\cite{10}. Therefore, it is essential to develop an accurate CT-MR synthesis model to support these critical applications ~\cite{8}, ~\cite{9}. The term "CT-MR image synthesis" is used in this paper to refer to the bidirectional translation between CT and MR images.

CycleGAN is an unsupervised machine learning technique designed for unpaired image-to-image translation tasks across different domains~\cite{5}. Unlike traditional supervised learning that relies on paired datasets, CycleGAN operates without the need for corresponding pairs of images. It achieves this through a Generative Adversarial Network (GAN) architecture consisting of a generator and a discriminator~\cite{49}. The innovation lies in its use of cycle-consistency loss, which ensures that translated images maintain content integrity when moving between source and target domains. Through adversarial training, CycleGAN learns to generate images that are indistinguishable from real ones in the target domain, enabling versatile cross-domain translations without paired data~\cite{50,18}.   

CT-MR image synthesis is a challenging task due to several issues. Firstly, obtaining CT and MR images separately is often time-consuming, costly, and burdensome to the patient. As a result, most available training datasets are unpaired. Secondly, the limited number of available training datasets for CT-MR image synthesis is a significant issue, with most datasets being small. This problem is critical because the number of training images is essential for generating realistic images. To overcome this issue, pre-trained models is a commonly used technique that can improve the GAN model's performance and generate more realistic images~\cite{14}. It enables the GAN model to leverage knowledge gained from training on other datasets, which can then be applied to the CT-MR image synthesis task. It can reduce the amount of time and computational resources required to train deep learning models, which is particularly advantageous in medical imaging applications that have time constraints and require real-time processing. Furthermore, pre-trained models have been shown to improve deep learning models' generalization capabilities, lowering the risk of overfitting and improving their ability to handle unknown data~\cite{5}. However, pre-trained models may have some limitations, such as the possibility that they may not be appropriate for the new task or that they may not have been trained on similar data, which can result in poor performance or over-fitting~\cite{15}. Additionally, they may have been trained on data that contains biases or errors, which can negatively impact the model's performance on the new task~\cite{16}. Despite these limitations, pre-trained models are a powerful technique that can improve model performance in medical imaging and other fields. It is important to assess the suitability of the pre-trained models for the new task and to use appropriate data and regularization techniques to avoid over-fitting and biases~\cite{17}. There have been several studies in recent years that have applied pre-trained models to MR-CT image translation ~\cite{10}. These studies have shown promising results, improving the performance and efficiency of MR-CT image translation models.

\begin{figure*}[!htbp]
\centering
\includegraphics[width=18 cm, height=18 cm]{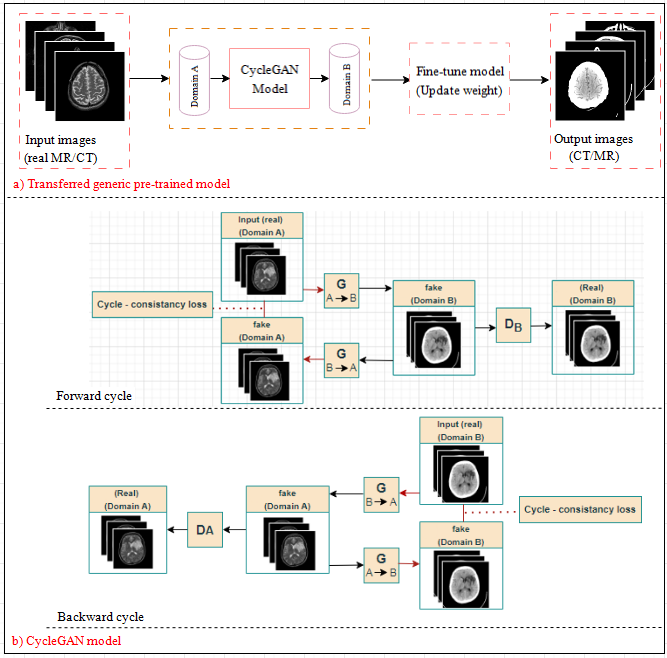}
\caption{Transfer learning from generic pre-trained models.}
\label{fig_1}
\end {figure*}

This work aims to explore the effect of applying transfer learning from generic pre-trained models on MR-to-CT image synthesis. This would assist in investigating the factors that contribute to the success of applying a pre-trained model in medical image translation tasks. To achieve this goal, a paired MR-CT dataset was used with 18 different pre-trained models. The image translation efficacy will be evaluated using various image quality metrics such as peak signal-to-noise ratio (PSNR), structural similarity index (SSIM), universal quality index (UQI), and visual information fidelity (VIF). The paper investigates the outcome of the different pre-tained models on the translated image characteristics. The framework applied in this work relies on the general CycleGAN approach~\cite{30}. Figure~\ref{fig_1} shows the applied architectures, including CycleGAN. The paper aims to highlight the importance of understanding generic pre-trained models when used across domains, particularly for improving MR-to-CT synthesis, which may lead to better development of trustworthy artificial intelligence tools in medical imaging. The main contributions of this work are:
\begin{enumerate}
\item Analyze the effect of transfer learning from generic pre-trained models on bi-directional MR-CT image translation. 

\item Evaluate the performance of various generic pre-trained models comprehensively using four widely used image quality metrics which highlights different aspects of the generated image.

\item Assess the clinical significance of the translated MR-CT images through a perceptual study by two experienced radiologists.

\item Investigate the effect of image structure and discontinuities of the pre-trained model on the performance of medical image translation.
\end{enumerate}

The rest of the paper is organized as follows: Section II presents a review of related work of image synthesis in the medical imaging domain. Section III provides coverage of image-to-image translation using GAN models. Section IV explains the proposed paired-unpaired unsupervised learning model with transfer learning. Section V illustrates experimental results and performance. Section VI analyses and interprets main findings. Finally, Section VII concludes and gives future directions. 

\section{Related work}
In the field of medical imaging there has been an amount of research dedicated to achieving accurate and efficient brain tumor detection ~\cite{58,59, 60}, classification and prediction of tumor grade ~\cite{61, 62}. A major advancement, in this area has been the use of pre-trained models especially in the context of translating MR images to CT images. The ultimate goal is to make the process of translating images more precise. In this section we will explore the studies that have utilized pre-trained models to address important tasks such, as detecting tumors classifying them accurately and determining their grade.

\subsection{Brain Tumor Detection}

Deep transfer learning was employed by Chelghoum et al.~\cite{25} to classify three types of brain tumors with contrast-enhanced magnetic resonance images (CE-MRI) and achieved a high classification accuracy of 98.71\%, demonstrating the reliability of transfer learning for small datasets. However, one potential disadvantage is the dependence on a specific dataset. The model's performance and reliability may be limited to the characteristics of that dataset, such as imaging protocols, patient demographics, and tumor variations. The high accuracy achieved on the current dataset does not guarantee similar results on different datasets. Arbane et al.~\cite{26} use transfer learning and various CNN architectures, including ResNet, Xception, and MobilNet-V2, for brain tumor classification from MRI images. The MobilNet-V2 architecture achieved the highest accuracy of 98.24\% and an F1-score of 98.42\%. The main drawback is the lack of in-depth understanding regarding the specific factors that contribute to MobilNet-V2's superior performance compared to other architectures. While the results are impressive, gaining insights into the underlying reasons for this performance disparity can aid in model optimization and generalization to different datasets. while Kaur et al.~\cite{27} replaced the last three layers of eight pre-trained CNNs for brain tumor classification using MRI images, achieving a classification accuracy of 96.05\%. While this transfer learning approach demonstrated effectiveness in the domain, one potential disadvantage is the limited exploration of alternative modifications to the network architecture. By solely replacing the last three layers, the study may not have fully optimized the network for brain tumor classification. Important features or patterns in the earlier layers might not have been fully utilized or adapted. Also, Saxena et al.~\cite{28} The study aimed to classify brain tumor cells as cancerous or non-cancerous using pre-trained VGG16, ResNet50, and Inceptionv3 networks. The researchers employed preprocessing and augmentation techniques to improve the datasets and achieved the highest accuracy of 95\% with the ResNet50 network. Considering alternative network architectures or fine-tuning the existing models could enhance accuracy and adaptability to the unique characteristics of brain tumor datasets. Çinar et al.~\cite{29} employed the ResNet50 architecture for tumor detection in MRI images. They replaced the last 5 layers of ResNet50 with 8 other layers and experimented with other pre-trained deep networks. The MRI dataset was split into 60\% for training and 40\% for testing, achieving an impressive accuracy of 97.2\%.  However, it is unclear whether the high accuracy is solely attributed to the specific dataset used or if it can be replicated across different datasets. Abu-Srhan et al. ~\cite{5} introduce a method called uagGAN which stands for Paired Unsupervised Attention Guided Generative Adversarial Network. This approach combines transfer learning with the goal of synthesizing brain MR and CT images, in both directions. The uagGAN model takes advantage of paired and unpaired datasets using a process of pre-training and retraining. Notably, attention masks are utilized to improve the accuracy and sharpness of the generated images. Additionally, the model incorporates transfer learning from a trained nonmedical model to enhance the learning process and improve image translation performance. The proposed methodology has shown results in evaluations as well as qualitative assessments by radiologists demonstrating its superior ability to generate accurate brain MR and CT images bi-directionally compared to existing image, to image translation techniques.

\subsection{Brain Tumors Classification}
In the context of brain tumor classification, a non-invasive MRI-based approach was introduced by Tandel et al.~\cite{20}. Their method is designed to differentiate between LGG and HGG tumors and relies on the amalgamation of five pre-trained CNNs via a majority voting mechanism, which significantly augments the classification accuracy. This approach exhibits enhanced accuracy when applied across three distinct datasets. Nevertheless, it's worth noting that the decision-making process underpinning this algorithm may present challenges in terms of comprehensibility and explication. Also, Rehman et al.~\cite{22} utilized three deep networks, AlexNet, VGG16, and GoogLeNet, with transfer learning, pre-processing, and data augmentation to classify brain tumors. The highest accuracy of 98.69\% was achieved by the VGG16 network. Training and fine-tuning multiple networks like AlexNet, VGG16, and GoogLeNet can be computationally intensive and time-consuming. while Deepak et al.~\cite{23} used transfer learning with a pre-trained deep neural network, GoogLeNet, to classify brain tumors, achieving a mean accuracy of 98\%. They only modified the last three layers of GoogLeNet and trained it with a SoftMax classifier. Additionally, they used GoogLeNet as a feature extractor for SVM and KNN classifiers. While this yielded high accuracy, it restricts the ability to adapt the model to dataset-specific characteristics or incorporate domain-specific knowledge. Swati et al.~\cite{24}  focused on utilizing deep learning for extracting features from medical images, specifically MRIs. They employed a pre-trained VGG19 model and applied block-wise fine-tuning transfer learning to adapt the learned features from general images to medical images. The study achieved an accuracy of 94.82\%. The VGG19 model was trained on a different dataset, likely containing different image types, potentially leading to suboptimal performance in capturing specific features and patterns present in medical images, particularly MRIs. Also, Hao et al.~\cite{21} proposed a transfer learning-based active learning framework that aims to reduce annotation costs while maintaining model stability and robustness for brain tumor classification. It achieves an AUC of 82.89\% on a balanced dataset of 82\%. The lower AUC value indicates that the model's ability to distinguish between positive and negative cases may not be optimal. Additionally, the balanced dataset of 82\% suggests a potential class imbalance issue, which could result in a bias towards the majority class and impact the model's performance on minority classes. 

\subsection{Tumor Grade Prediction} 
 In the domain of glioma tumor grade prediction from brain MR images, Yang et al.~cite{19} applied deep learning models, including AlexNet and GoogLeNet. Their investigation yielded compelling results, as both models demonstrated a remarkable level of accuracy in tumor grade identification. These findings underscore the promising role of deep learning in advancing medical image analysis. However, it is imperative to recognize that the study's assessment was limited to a singular dataset, prompting considerations regarding the broader applicability of these results across diverse datasets and clinical scenarios. In the same way, Tenghongsakul et al. ~\cite{44}  compared different deep transfer learning methods (InceptionResNet-V2, ResNet50, MobileNet-V2, and VGG16) for brain tumor prediction using a public MRI dataset. Image enhancement technique CLAHE was applied to improve image quality. The suggested method achieved a remarkable prediction accuracy of up to 100\%. while Khan et al.~\cite{43} investigate the application of CNNs for image-based brain cancer prediction to enhance treatment reliability. Using transfer learning with pre-trained VGG19 and MobileNetV2 models, the research focuses on improving accuracy. The experiment achieved 97\% accuracy with MobileNetV2 and 91\% with VGG19.Table ~\ref{T1} presents a summary of the studies discussed above and provides a quick reference guide for each study.

\begin{table*}[!htbp]
\caption{\label{T1} A Review of Related Works}
\centering
\begin{tabular}{l|c|c|c|c|c|c|}
\cline{2-7}
                                                                            & \textbf{Network backbone}                                                                                                                      & \textbf{Training type}                                                                                         & \multicolumn{1}{l|}{\textbf{Unsupervised}} & \textbf{Bi-directional} & \textbf{Dataset}                                                                                                                                      & \textbf{Metrics}                                                                                                                                                                   \\ \hline
\multicolumn{1}{|l|}{Yang et al. ~\cite{19}}          & AlexNet and GoogLeNet                                                                                                                          & \begin{tabular}[c]{@{}c@{}}·   Cross-validation \\ (5-fold CV)\end{tabular}                                    &                                  &                         & \begin{tabular}[c]{@{}c@{}}Brain   MR Images, \\ 113 Glioma Patients\end{tabular}                                                                     & Accuracy,   AUC                                                                                                                                                                    \\ \hline
\multicolumn{1}{|l|}{Tenghongsakul et al. ~\cite{44}} & \begin{tabular}[c]{@{}c@{}}InceptionResNet-V2,\\    ResNet50, \\ MobileNet-V2,\\  VGG16\end{tabular}                                           & Not specified                                                                                                  & $\checkmark$                               &                         & \begin{tabular}[c]{@{}c@{}}Brian   Tumor Dataset\\  from kaggle\end{tabular}                                                                          & \begin{tabular}[c]{@{}c@{}}Accuracy,   Precesion, \\ Recall, F1- score\end{tabular}                                                                                                \\ \hline
\multicolumn{1}{|l|}{Khan et al. ~\cite{43}}          & VGG19 , MobileNetV2                                                                                                                            & Transfer learning                                                                                              & $\checkmark$                               & $\checkmark$            & \begin{tabular}[c]{@{}c@{}}Brian   Tumor Dataset \\ from kaggle\end{tabular}                                                                          & Accuracy,   F1                                                                                                                                                                     \\ \hline
\multicolumn{1}{|l|}{Tandel et al.~\cite{20}}         & \begin{tabular}[c]{@{}c@{}}AlexNet,   VGG16, \\ ResNet18, GoogleNet, \\ ResNet50\end{tabular}                                                  & \begin{tabular}[c]{@{}c@{}}Not specified \end{tabular}                                     &                                            & $\checkmark$            & \begin{tabular}[c]{@{}c@{}}Three different datasets: \\ RSM, SSM, and WBM data\end{tabular}                                                           & Accuracy                                                                                                                                                                           \\ \hline
\multicolumn{1}{|l|}{Rehman et al.~\cite{22}}         & \begin{tabular}[c]{@{}c@{}}AlexNet, GoogLeNet, \\ VGGNet\end{tabular}                                                                          & \begin{tabular}[c]{@{}c@{}}Transfer Learning \\ (Fine-Tune and Freeze)\end{tabular}                            &                                            & $\checkmark$            & \begin{tabular}[c]{@{}c@{}}Brain   tumor MRI \\ slices from Figshare\end{tabular}                                                                     & Accuracy                                                                                                                                                                           \\ \hline
\multicolumn{1}{|l|}{Deepak et al. ~\cite{23}}        & CNN with SVM                                                                                                                                   & Fivefold Cross-Validation                                                                                      & $\checkmark$                               &                         & \begin{tabular}[c]{@{}c@{}}Figshare open dataset \\ containing MRI images\end{tabular}                                                                & Accuracy                                                                                                                                                                           \\ \hline
\multicolumn{1}{|l|}{Swati et al. ~\cite{24}}         & \begin{tabular}[c]{@{}c@{}}Pre-trained  \\ deep CNN model\end{tabular}                                                                         & \begin{tabular}[c]{@{}c@{}}Transfer Learning \\ (Block-Wise Fine-Tuning)\end{tabular}                          & $\checkmark$                               & $\checkmark$            & \begin{tabular}[c]{@{}c@{}}T1-weighted \\ contrast-enhanced \\ magnetic resonance \\ images (CE-MRI)\\  benchmark dataset\end{tabular}                & Accuracy                                                                                                                                                                           \\ \hline
\multicolumn{1}{|l|}{Hao et al. ~\cite{21}}           & \begin{tabular}[c]{@{}c@{}}Pre-trained model \\ as a backbone \\ (2D slice-based \\ approach)\end{tabular}                                     & \begin{tabular}[c]{@{}c@{}}Transfer Learning \\ (Active Learning)\end{tabular}                                 &                                            &                         & \begin{tabular}[c]{@{}c@{}}MRI  training dataset \\ of 203 patients \\ and a validation \\ dataset of 66 patients\end{tabular}                        & AUC                                                                                                                                                                                \\ \hline
\multicolumn{1}{|l|}{Chelghoum et al. ~\cite{25}}     & \begin{tabular}[c]{@{}c@{}}Nine deep pre-trained \\ CNN architectures\end{tabular}                                                             & \begin{tabular}[c]{@{}c@{}}Cross-validation \\ (5-fold CV)\end{tabular}                                        & $\checkmark$                               & $\checkmark$            & \begin{tabular}[c]{@{}c@{}}Brain CE-MRI \\ benchmark dataset\end{tabular}                                                                             & Accuracy                                                                                                                                                                           \\ \hline
\multicolumn{1}{|l|}{Arbane et al. ~\cite{26}}        & \begin{tabular}[c]{@{}c@{}}ResNet, Xception, \\ MobilNet-V2\end{tabular}                                                                       & Not specified                                                                                                  &                                            &                         & MRI   Images                                                                                                                                          & Accuracy   and F1-score                                                                                                                                                            \\ \hline
\multicolumn{1}{|l|}{Kaur et al. ~\cite{27}}          & \begin{tabular}[c]{@{}c@{}}AlexNet,   Resnet50, \\ GoogLeNet, VGG-16, \\ Resnet101, VGG-19, \\ Inceptionv3,  \\ InceptionResNetV2\end{tabular} & \begin{tabular}[c]{@{}c@{}}Transfer Learning using \\ Pre-trained DCNNs\end{tabular}                            &     $\checkmark$                                        &                         & \begin{tabular}[c]{@{}c@{}}Harvard, Clinical, \\ and Benchmark \\ Figshare Repository\end{tabular}                                                    & \begin{tabular}[c]{@{}c@{}}Accuracy,   Sensitivity, \\ Specificity, Precision, \\ False Positive Rate, \\ Error, F1-score, AUC,\\  Mathew Correlation \\ Coefficient.\end{tabular} \\ \hline
\multicolumn{1}{|l|}{Saxena et al. ~\cite{28}}        & \begin{tabular}[c]{@{}c@{}}Resnet-50, VGG-16, \\ Inception-V3\end{tabular}                                                                     & \begin{tabular}[c]{@{}c@{}}CNN-based\\  Transfer Learning\end{tabular}                                         & $\checkmark$                               & $\checkmark$            & \begin{tabular}[c]{@{}c@{}}Brain MRI images,\\  which contains \\ a total of 253 images\end{tabular}                                                  & Accuracy                                                                                                                                                                           \\ \hline
\multicolumn{1}{|l|}{Çinar et   al.~\cite{29}}        & \begin{tabular}[c]{@{}c@{}}Resnet50, Alexnet, \\ Densenet201, \\ InceptionV3, \\ Googlenet\end{tabular}                                        & \begin{tabular}[c]{@{}c@{}}Fivefold \\ Cross-Validation\end{tabular}                                           &                                            &                         & MRI   Images                                                                                                                                          & Accuracy                                                                                                                                                                           \\ \hline
\multicolumn{1}{|l|}{Abu-Srhan et al. ~\cite{5}}      & \begin{tabular}[c]{@{}c@{}}Paired-unpaired \\ Unsupervised \\ Attention Guided\\  GAN\end{tabular}                                             & \begin{tabular}[c]{@{}c@{}}Pre-training \\ with Paired \\ Data and Retraining \\ on Unpaired Data\end{tabular} & $\checkmark$                               & $\checkmark$            & \begin{tabular}[c]{@{}c@{}}The dataset by \\ Han et al. {[}13{]} \\ contained 367 \\ paired MR-CT \\ brain images \\ from 18   patients.\end{tabular} & PSNR,   SSIM, UQI, VIF                                                                                                                                                             \\ \hline
\end{tabular}
\end{table*}

In this work, we investigate 18 different pre-trained models to translate 400 brain MRI images to CT images and vice versa. The goal of the study was to explore the ability of these models to perform image translation between different modalities and generate images that are comparable to real images. Models are chosen to be comprehensive and based on their performance in previous studies.

\section{Materials and Methods}
Pre-trained models that are used in this work are compatible with the CycleGAN framework~\cite{30} and cover diverse domains such as natural scenes, animals to urban and satellite images.

\subsection{Cycle GAN}

\begin{figure}[!htbp]
\centering
\includegraphics[width=4cm, height=3cm]{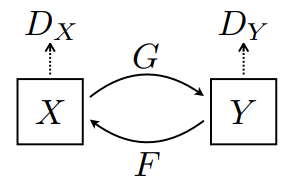}
\caption{CycleGAN Generators and Discriminators}
\label{fig_2}
\end{figure}

CycleGAN is a type of Generative Adversarial Network (GAN) that can translate one image type to another. Its primary objective is to learn how to map between two image domains without the use of paired training data, in an unsupervised manner~\cite{30}. CycleGAN consists of two generators, G and F, and two discriminators, Dx and Dy as shown in Figure ~\ref{fig_2} ~\cite{30}. The generators are responsible for changing images from one domain to another, while the discriminators learn to differentiate between genuine images and fake ones generated by the generators. The diagram in Figure~\ref{fig_1} (b) illustrates the architecture of CycleGAN~\cite{31}.The network includes both forward and backward cycles. In the forward cycle, there are two generators and one discriminator. Generator-CT is responsible for learning the mapping from MR to CT to produce artificial CT images. Discriminator-CT, on the other hand, is trained to distinguish between real and synthetic CT images and to encourage Generator-CT to produce realistic images. Generator-MR, which learns the CT-to-MR mapping, helps to improve the ability of Generator-CT to synthesize accurate CT images. The Cycle consistency loss is used to ensure that the synthesized CT image can be converted back to the original MR image. In the backward cycle, the structure is identical to the forward cycle, except that CT images are used as input instead of MR images~\cite{31}.
The loss function for the CycleGAN includes two components: the adversarial loss and the cycle consistency loss.The adversarial loss is given by \ref{e1}: In ~\cite{31}:
\begin{equation} \label{e1}
\begin{split}
L_{GAN}(G, D_{Y}, X, Y)= E_{y\sim p_{data(y)}}\left [  \log D_{Y}\left (  y\right )\right ] \\
+ E_{x\sim p_{data(x)}}\left [  \log \left (  1-D_{Y}G \left ( x \right )\right )\right ]
\end{split}
\end{equation}

Where $G$ tries to generate images $G\left( X \right)$ that look similar to images from domain $Y$, while $D_{Y}$ aims to distinguish between translated samples $G\left( X \right)$ and real samples $y$. $G$ aims to minimize this objective against an adversary $D$ that tries to maximize it, i.e., $min_{G}max_{D_{Y}}L_{GAN}\left( G, D_{Y}, X, Y\right)$, a similar adversarial loss for the mapping function is introduced $F: Y \to X$ and its discriminator $D_{X}$ as well: i.e., $min_{F}max_{D_{X}}L_{GAN}\left( G, D_{Y}, Y, X\right)$. The adversarial loss encourages the generator $G$ to generate images that are indistinguishable from real images from domain $Y$, and the discriminator $D_{y}$ to correctly distinguish between real and fake images. The cycle consistency loss is given by \ref{e2} ~\cite{30}: 
\begin{equation} \label{e2}
\begin{split}
 L_{cyc}\left ( G, F \right ) = E_{x\sim p_{data \left ( x \right )}}\left [ \left \| F\left ( G\left ( X \right ) \right ) - x \right \|_{1} \right] \\ + E_{y\sim p_{data \left ( y \right )}}\left [ \left \| G\left ( F\left ( y \right ) \right ) - y \right \|_{1} \right] 
 \end{split}
\end{equation}

For each image $x$ from domain $X$, the image translation cycle should be able to bring $x$ back to the original image, i.e., $x\to G\left( x \right)\to F\left( G\left(  x\right) \right) \approx x$. 

Similarly, for each image $y$ from domain $Y$, $G $and $F$ should also satisfy backward cycle consistency: $y\to F\left( y \right)\to G\left( F\left(  y\right) \right) \approx y$.

The cycle consistency loss encourages the generator $F(G(x))$ to be like $x$ and $G(F(y))$ to be like $y$. The objective is to learn mapping functions between the two domains $X$ and $Y$ using training samples with the labels $\mathrm{\left\{ x_{i} \right\}}_{i=1}^{N}$, where $x_{i}\in X$ and $\mathrm{\left\{ y_{j} \right\}}_{j=1}^{N}$, where $y_{j}\in Y$. 

The data distribution is represented as $x\sim p_{data(x)}$ and $y\sim p_{data(y)}$. This model includes two mappings $G: X \to Y$ and $F: Y\to X$.

Additionally, The two adversarial discriminators $D_{X}$ and $D_{Y}$ are introduced, where $D_{X}$ aims to distinguish between images ${x}$ and translated images ${F(y)}$; in the same way, $D_{Y}$ aims to discriminate between ${y}$ and ${G(x)}$ ~\cite{30}. 

\subsection{Transfer learning and Fine-tuning}
Transfer learning is a technique in machine learning that involves using a model trained on one task to enhance the performance of a model trained on a different but related task. The primary objective of transfer learning is to make use of the knowledge and experience gained from one problem to improve the model's performance on another ~\cite{33}. This involves utilizing a pre-trained model as a starting point and fine-tuning it for the new task using a smaller dataset~\cite{35}. Transfer learning has been extensively employed in various fields, including computer vision, natural language processing, and speech recognition. It is particularly useful in labelling pre-trained labelled data for the target task is scarce or when the target task is related to the source task~\cite{30}.

In the medical field, transfer learning is a potent technique that can be used to translate images ~\cite{34}. Medical image translation is the process of converting an image from one modality to another, such as from an MR image to a CT image or vice versa. The primary objective of medical image translation is to generate images that resemble the originals and contain the same information. Transfer learning can be utilized in medical image translation to enhance the model's performance on a smaller dataset specific to the medical domain by leveraging the knowledge obtained from a more extensive dataset of images ~\cite{36}. For instance, a model trained on a vast dataset of natural images can be fine-tuned on a smaller dataset of medical images to improve its accuracy on the medical image translation task. According to ~\cite{37}, transfer learning is advantageous in medical image translation because it significantly reduces the amount of labelled data needed for training. This is especially useful in medical imaging, where obtaining labelled data can be challenging and expensive. Additionally, transfer learning can aid models in performing better on unseen data by improving their generalization ability. This is critical in medical imaging, where the model must function effectively on a broad range of patients and conditions ~\cite{35}.

In transfer learning ~\cite{33}, we define a domain as $D = \left\{ X, P\left( x \right) \right\}$ , where $X$ is the feature space with $X = \left\{ x_{1},\cdots ,x_{n} \right\}\subset X$ and $P(X)$ is a marginal probability distribution. For example, $X$ could include all possible images derived from a particular MRI protocol, acquisition parameters, and scanner hardware, and $P(X)$ could depend on, for instance, subject groups, such as adolescents or elderly people. 
P(X) for MR to CT translation represents the probability distribution of intensities or features within the MR images, considered independently of their corresponding CT counterparts. The value can be obtained by the integral sums over all possible values of y, representing the marginalization of the joint distribution over the CT dimension to obtain the marginal distribution of MR. Tasks comprise a label space Y and a decision function $f$, i.e., $T= \left\{ Y, f \right\}$. The decision function is to be learned from the training data $\left( X, Y \right)$. Tasks in MR brain imaging can be, for instance, the survival rate prediction of cancer patients, where $f$ is the function that predicts the survival rate and $Y$ is the set of all possible outcomes. Given a source domain $D_{S}$ and task $T_{S}$ and a target domain $D_{T}$ and task $T_{T}$, transfer learning re-utilizes the knowledge acquired in $D_{S}$ and $T_{S}$ to improve the generalization of $f_{T}$ in $D_{T}$~\cite{38}. Importantly, $D_{S}$ must be related to $D_{T}$ and $T_{S}$ must be related to $T_{T}$ ~\cite{38}; otherwise, transfer learning can worsen the accuracy in the target domain. This phenomenon, called negative transfer, has been recently formalized in ~\cite{35} and studied in the context of MR brain imaging ~\cite{31}.

In our approach, inspired by the network architectures outlined in ~\cite{30}, we augment each pre-trained model by incorporating two new fully-connected layers. These layers are initialized with random weights and are accompanied by a Softmax activation function. Then, we freeze the weights of the remaining layers in the network to prevent them from being updated during the training process. This ensures that the learned features in these layers, derived from the pre-trained model, remain intact while we focus on fine-tuning the added layers to suit our new data.

The fine-tuning of parameters, particularly $\lambda$, is crucial and depends on the specific characteristics of the pre-trained model and we set according to its category as the following: for "Artistic Style Transfer" $\lambda = 9 $;  "Animal Images", "Natural Landscape Images" and "Photography" $\lambda = 10 $; "Satellite and Map Images" and "Urban Scenes" $\lambda = 11$; Experiments have demonstrated the sensitivity of the results to changes in the $\lambda$ value. Other parameters are set as follows: the batch size is 2, FC1 has 256 neurons, FC2 has 256 neurons (adjusted according to the dimensions of the CT and MR images), training is conducted for 200 epochs, and the activation function employed is Softmax. These hyper-parameters were carefully selected based on the outcomes of our experiments, allowing us to tailor the model's performance to meet our objectives and accommodate the dataset's properties. During training, We keep the same learning rate for the first 100 epochs and linearly decay the rate to zero over the next 100 epochs This approach enables smoother convergence and aids in fine-tuning the model effectively. Algorithm 1 illustrates the bidirectional CT-MR translation using transfer learning from generic pre-trained models.

\begin{algorithm}
\caption{Bidirectional CT-MR translation using transfer learning from pre-trained models}
\begin{algorithmic}
\label{A1}
\Require {Generic pre-trained model, dataset (X $\leftarrow$ CT, Y  $\leftarrow$ MR)}
\Ensure {Fine-tuned model}
\For{each pre-trained model}
        \State Load the pre-trained Model
        \State Initialize the new model with the pre-trained weights
        \State Finetune\textunderscore model:
        \State (FC1 = 256; FC2 = 256;learning rate = 0.001; 
        \State activation = "softmax"; epoch = 200; batch = 2)
        \\
        \State Add\textunderscore layer ($FC1$, activation)
        \State Add\textunderscore layer ($FC2$, activation)
        \State Freeze pre-existing layers
        \State Set learning rate 
        \State Loss Function: 
          \begin{equation} 
          \begin{split}
               L(G,F,D_{X},D_{Y}) = L_{GAN}(G,D_{Y},X,Y) \\ + L_{GAN} (F,D_{X}, Y,X) + \lambda L_{cyc}(G,F)
         \end{split}
         \end{equation}
        \State Build a new model
\While{not converged}
    \For{each epoch}
        \For{each batch of images from the target domain}
            \State Train new model ($X$, $Y$, $D_{X}$, $D_{Y}$).
        \EndFor
    \EndFor
    \State Evaluate new model ($X$, $Y$, $D_{X}$, $D_{Y}$).
\EndWhile
    \EndFor
\end{algorithmic}
\end{algorithm}

\subsection{pre-trained model} 
Generic pre-trained models refer to models that are trained on datasets that are not specific to the medical field. These models can be used as a starting point for medical image analysis tasks such as segmentation, classification, and detection. In this study, we investigate the effect of using non-medical pre-trained models on MR-to-CT synthesis ~\cite{39}. 

We used 18 generic pre-trained models for the task of medical bi-directional MR-CT translation. These models were selected from a range of publicly available models, for different types of images, including Artistic Style Transfer, Animal Images, Natural Landscape Images, Photography, Satellite and Map Images, and Urban Scenes, which can be used as a basis for developing models for MR to CT translation. These pre-trained models are trained on large-scale datasets of high-quality images and can learn high-level features that can be transferred to other domains, such as medical imaging. By leveraging pre-trained models, researchers can save significant amounts of time and computational resources and achieve better results than by training models from scratch. For example, the style transfer pre-trained models can learn the underlying style of an image and apply it to another image, which can be useful in developing models for translating MR to CT images with similar styles. Similarly, for animal and natural landscape images, pre-trained models can learn features specific to those domains, which can be used in developing models for medical imaging.

Also, pre-trained models that were trained on satellite and map images can be used to make models for translating images from MRI to CT that were taken with imaging methods like computed tomography (CT) and magnetic resonance imaging (MRI). Satellite and map images often show the edges and shapes of buildings, roads, and other structures, just like medical images do. Also, models that have already been trained for urban scenes can learn details about the buildings and objects that are common in cities. These details can be used to create models for medical imaging of organs and tissues in the body that have similar details.

The models that were used fall into the following categories and figure ~\ref{fig_pre-trained model} shows an example of these categories:
     \begin{figure*}[!htbp]
    \centering
    \includegraphics[width=15cm, height=18cm]{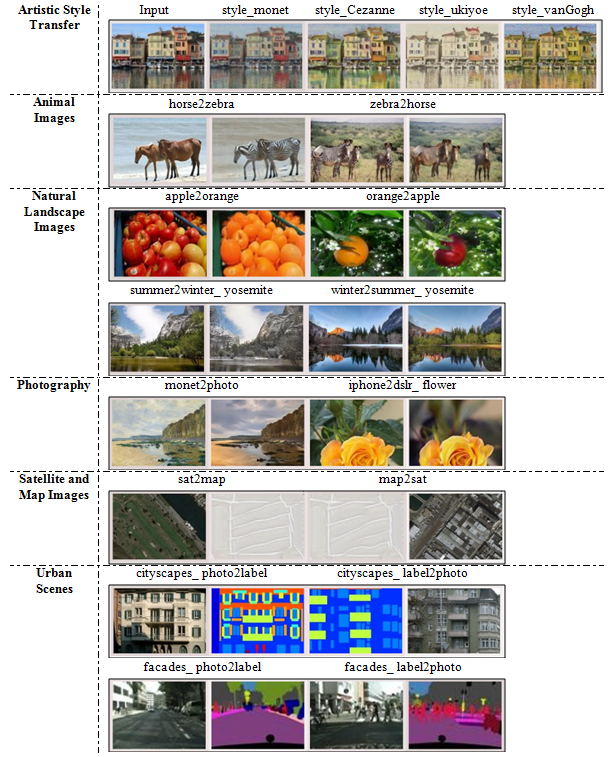}
    \caption{Examples of generic pre-trained models.}
    \label{fig_pre-trained model}
    \end{figure*}

    \begin{enumerate}
  \item \textbf{Artistic Style Transfer:}  Style transfer is a technique that utilizes pre-trained models to apply the artistic style of paintings to other images. Models like style\textunderscore monet, style\textunderscore Cezanne, style\textunderscore ukiyoe, and style\textunderscore vanGogh are trained on various artists' paintings. The method optimizes both content and style, creating new images with unique characteristics. It uses content and style loss functions, minimizing them to combine target image content with a reference image style. This technique finds applications in image synthesis, style transfer, and domain adaptation, making it valuable for artists, designers, and social media influencers seeking visually appealing and stylized images.
  
  \item \textbf{Animal Images:} Pre-trained models use animal images for object detection, classification, and segmentation. Models like horse2zebra and zebra2horse are trained on labelled datasets to generate images resembling horses or zebras. Variability in animal appearance is crucial. Models are optimized with backpropagation and stochastic gradient descent. Applications include wildlife conservation, ecology, agriculture, and entertainment. Accurate animal detection aids in preserving endangered species and promoting biodiversity.
  
 \item \textbf{Natural Landscape Images:} Pre-trained image-to-image translation models transform landscapes between different seasons while preserving key features. Examples include apple2orange, orange2apple, summer2winter\textunderscore yosemite, and winter2summer\textunderscore yosemite. Trained on natural Yosemite landscape images, the models depict forests, mountains, rivers, and beaches. Classifying and recognizing objects in such complex landscapes can be challenging. Optimization is done with datasets like Places2 and Open Images. Applications include object recognition, semantic segmentation, scene recognition, and more, benefiting ecology, tourism, and climate change research for analyzing environmental changes and their impact on ecosystems.
 
\item \textbf{Photography:} Image-to-image translation involves using pre-trained models like monet2photo and iphone2dslr\textunderscore flower for transforming images from one style to another. These models are trained on datasets of Monet paintings and real photographs to create images with similar appearance. They exhibit high resolution, rich texture, and detailed visual features, including lighting, shadows, and color. Object detection, recognition, and segmentation are incorporated to classify and locate objects accurately. Optimization is done with large datasets like ImageNet, COCO, Pascal VOC, and Open Images. These versatile models find applications in surveillance, healthcare, and education, such as disease diagnosis in medical imaging, realistic simulations in education, and object detection in surveillance scenarios.

\item \textbf{Satellite and Map Images:} In computer vision, the analysis of aerial or satellite images of the Earth's surface is done. Pre-trained models like sat2map and map2sat perform image-to-image translation of satellite and map images with high spatial resolution, enabling the identification of small-scale surface features. These models incorporate land cover classification, object detection, and change detection for environmental analysis. Optimized through transfer learning, they find applications in urban planning, agriculture, natural resource management, and disaster response. For instance, they aid in urban growth analysis, crop monitoring, land use change detection, and disaster-affected area identification for rapid response and relief efforts.
    
\item \textbf{Urban Scenes:} Pre-trained models like cityscapes\textunderscore photo2label, cityscapes\textunderscore label2photo, facades\textunderscore photo2label, and facades\textunderscore label2photo are designed for image-to-image translation between urban scene images and their corresponding semantic label maps. Trained on datasets of urban scene images and semantic label maps, these models facilitate object detection, segmentation, and classification. Urban scenes pose challenges with complexity, high density, dynamic elements, and varying lighting conditions, making traditional computer vision techniques less effective. Convolutional neural networks (CNNs) used in these models enable accurate and efficient performance. Optimization with pre-trained models enhances their effectiveness, especially with limited training data. Applications include urban planning, traffic management, and security, automating urban scene analysis with machine learning and computer vision techniques. Examples include traffic flow analysis, vehicle and pedestrian detection and tracking, and public safety and security monitoring in urban areas.
\end{enumerate} \leavevmode

The pre-trained models mentioned in the provided description utilize the same network architecture for both the generator and discriminator components. However, the key point of differentiation lies in the training dataset used for each model. The network architecture is adapted from ~\cite{30}. 

Generator architectures:
This architecture employs six residual blocks for training images of size 128 × 128, while nine residual blocks are used for higher-resolution images such as 256 × 256 or beyond. The architecture comprises various layers denoted by specific notations: c7s1-k represents a 7 × 7 Convolution-Instance Norm ReLU layer with k filters and stride 1, dk denotes a 3 × 3 Convolution-Instance Norm-ReLU layer with k filters and stride 2, and Rk signifies a residual block containing two 3 × 3 convolutional layers with the same number of filters. Additionally, uk denotes a 3 × 3 fractional-strided Convolution-Instance Norm-ReLU layer with k filters and stride $\frac {1} {2}$. To mitigate artifacts, reflection padding is employed. 

For a generator network with six residual blocks, the architecture consists of: c7s1-64, d128, d256, followed by six R256 blocks, u128, u64, and finally c7s1-3. 

Meanwhile, a generator with nine residual blocks includes c7s1-64, d128, d256, followed by nine R256 blocks, u128, u64, and c7s1-3.

Discriminator architectures:
The discriminator architecture comprises layers denoted by Ck, representing a 4 × 4 Convolution-InstanceNorm-LeakyReLU layer with k filters and stride 2. Notably, InstanceNorm is excluded for the first C64 layer, and leaky ReLUs with a slope of 0.2 are employed. The discriminator architecture sequentially follows a structure of C64-C128-C256-C512, with a final convolutional layer producing a 1-dimensional output.

The Table ~\ref{T6} encapsulates crucial details regarding the images utilized in the training process for each pre-trained model ~\cite{30}. Each entry delineates the specific dataset employed, offering insights into the diversity and scale of image data utilized in the training regimen. These datasets serve as the foundational bedrock upon which the pre-trained models are built, encompassing a spectrum of visual information necessary for robust and comprehensive training. The effectiveness of these models in MR to CT translation tasks is then evaluated. By scrutinizing the provided dataset information, it is possible to determine which pre-trained model shows the highest efficiency at this specific translation task.

\begin{table*}[!htbp]
\caption{\label{T6} Datasets the Generic Pre-Trained Models trained on}
\centering

\begin{tabular}{|c|c|c|c|c|l|}
\hline
{\color[HTML]{000000} \textbf{Category}}                                                & {\color[HTML]{000000} \textbf{Pre-trained model}}                                     & {\color[HTML]{000000} \textbf{Photo type}}                & {\color[HTML]{000000} \textbf{\begin{tabular}[c]{@{}c@{}}Numbers of\\ training images\end{tabular}}} & {\color[HTML]{000000} \textbf{Image size}}         & \multicolumn{1}{c|}{{\color[HTML]{000000} \textbf{Source of images}}}                                                                            \\ \hline
{\color[HTML]{000000} }                                                                 & {\color[HTML]{000000} Style\_ monet}                                                  & {\color[HTML]{000000} monet}                              & {\color[HTML]{000000} 1074}                                                                          & {\color[HTML]{000000} }                            & {\color[HTML]{000000} }                                                                                                                          \\ \cline{2-4}
{\color[HTML]{000000} }                                                                 & {\color[HTML]{000000} Style\_ Cezanne}                                                & {\color[HTML]{000000} cezanne}                            & {\color[HTML]{000000} 584}                                                                           & {\color[HTML]{000000} }                            & {\color[HTML]{000000} }                                                                                                                          \\ \cline{2-4}
{\color[HTML]{000000} }                                                                 & {\color[HTML]{000000} Style\_ ukiyoe}                                                 & {\color[HTML]{000000} ukiyoe}                             & {\color[HTML]{000000} 1433}                                                                          & {\color[HTML]{000000} }                            & {\color[HTML]{000000} }                                                                                                                          \\ \cline{2-4}
\multirow{-4}{*}{{\color[HTML]{000000} Artistic Style Transfer}}                       & {\color[HTML]{000000} Style\_ vanGogh}                                                & {\color[HTML]{000000} vanGogh}                            & {\color[HTML]{000000} 401}                                                                           & \multirow{-4}{*}{{\color[HTML]{000000} 256 × 256}}   & \multirow{-4}{*}{{\color[HTML]{000000} \begin{tabular}[c]{@{}l@{}}Flickr API using the \\ "landscape" tag.\end{tabular}}}        \\ \hline
{\color[HTML]{000000} }                                                                 & {\color[HTML]{000000} }                                                               & {\color[HTML]{000000} zebra}                              & {\color[HTML]{000000} 1177}                                                                          & {\color[HTML]{000000} }                            & {\color[HTML]{000000} }                                                                                                                          \\ \cline{3-4}
\multirow{-2}{*}{{\color[HTML]{000000} Animal Images}}                                  & \multirow{-2}{*}{{\color[HTML]{000000} horse $\leftrightarrow$ zebra}}                                & {\color[HTML]{000000} horse}                              & {\color[HTML]{000000} 939}                                                                           & \multirow{-2}{*}{{\color[HTML]{000000} 256 × 256}}   & \multirow{-2}{*}{{\color[HTML]{000000} ImageNet ~\cite{53}}}                                                                                        \\ \hline
{\color[HTML]{000000} }                                                                 & {\color[HTML]{000000} }                                                               & {\color[HTML]{000000} apple}                              & {\color[HTML]{000000} 996}                                                                           & {\color[HTML]{000000} }                            & {\color[HTML]{000000} }                                                                                                                          \\ \cline{3-4}
{\color[HTML]{000000} }                                                                 & \multirow{-2}{*}{{\color[HTML]{000000} apple $\leftrightarrow$ orange}}                               & {\color[HTML]{000000} orange}                             & {\color[HTML]{000000} 1020}                                                                          & \multirow{-2}{*}{{\color[HTML]{000000} 256 × 256}} & \multirow{-2}{*}{{\color[HTML]{000000} ImageNet ~\cite{53}}}                                                                                        \\ \cline{2-6} 
{\color[HTML]{000000} }                                                                 & {\color[HTML]{000000} }                                                               & {\color[HTML]{000000} summer}                             & {\color[HTML]{000000} 1273}                                                                          & {\color[HTML]{000000} }                            & {\color[HTML]{000000} }                                                                                                                          \\ \cline{3-4}
\multirow{-4}{*}{{\color[HTML]{000000} Natural Landscape Images}}                       & \multirow{-2}{*}{{\color[HTML]{000000} summer $\leftrightarrow$ winter}}                                & {\color[HTML]{000000} winter}                             & {\color[HTML]{000000} 854}                                                                           & \multirow{-2}{*}{{\color[HTML]{000000} 256 × 256}} & \multirow{-2}{*}{{\color[HTML]{000000} \begin{tabular}[c]{@{}l@{}}Flickr API with the\\  tag "yosemite". \end{tabular}}} \\ \hline
{\color[HTML]{000000} }                                                                 & {\color[HTML]{000000} }                                                               & {\color[HTML]{000000} monet}                              & {\color[HTML]{000000} 1074}                                                                          & {\color[HTML]{000000} }                            & {\color[HTML]{000000} }                                                                                                                          \\ \cline{3-4}
{\color[HTML]{000000} }                                                                 & \multirow{-2}{*}{{\color[HTML]{000000} monet $\leftrightarrow$ photo}}                                  & {\color[HTML]{000000} photo}                              & {\color[HTML]{000000} 6853}                                                                          & \multirow{-2}{*}{{\color[HTML]{000000} 256 × 256}} & \multirow{-2}{*}{{\color[HTML]{000000} \begin{tabular}[c]{@{}l@{}}Flickr API with the\\  tag "monet".\end{tabular}}}                                \\ \cline{2-6} 
{\color[HTML]{000000} }                                                                 & {\color[HTML]{000000} }                                                               & {\color[HTML]{000000} iphone\_flower}                     & {\color[HTML]{000000} 1813}                                                                          & {\color[HTML]{000000} }                            & {\color[HTML]{000000} }                                                                                                                          \\ \cline{3-4}
\multirow{-4}{*}{{\color[HTML]{000000} Photography}}                                    & \multirow{-2}{*}{{\color[HTML]{000000} iphone $\leftrightarrow$ dslr flower}}                           & \multicolumn{1}{l|}{{\color[HTML]{000000} dslr \_flower}} & {\color[HTML]{000000} 3326}                                                                          & \multirow{-2}{*}{{\color[HTML]{000000} 256 × 256}} & \multirow{-2}{*}{{\color[HTML]{000000} \begin{tabular}[c]{@{}l@{}}Flickr API with the\\  tag "flower"\end{tabular}}}                       \\ \hline
\multicolumn{1}{|l|}{{\color[HTML]{000000} }}                                           & {\color[HTML]{000000} }                                                               & {\color[HTML]{000000} sat}                                & {\color[HTML]{000000} 400}                                                                           & {\color[HTML]{000000} }                            & {\color[HTML]{000000} }                                                                                                                          \\ \cline{3-4}
\multicolumn{1}{|l|}{\multirow{-2}{*}{{\color[HTML]{000000} Satellite and Map Images}}} & \multirow{-2}{*}{{\color[HTML]{000000} sat $\leftrightarrow$ map}}                                    & {\color[HTML]{000000} map}                                & {\color[HTML]{000000} 400}                                                                           & \multirow{-2}{*}{{\color[HTML]{000000} 256 × 256}} & \multirow{-2}{*}{{\color[HTML]{000000} Google Maps ~\cite{54}}}                                                                                    \\ \hline
{\color[HTML]{000000} }                                                                 & \multicolumn{1}{l|}{{\color[HTML]{000000} }}                                          & {\color[HTML]{000000} cityscapes}                         & {\color[HTML]{000000} 2975}                                                                          & {\color[HTML]{000000} }                            & {\color[HTML]{000000} }                                                                                                                          \\ \cline{3-4}
{\color[HTML]{000000} }                                                                 & \multicolumn{1}{l|}{\multirow{-2}{*}{{\color[HTML]{000000} Cityscapes\_photo $\leftrightarrow$ 
 label}}} & {\color[HTML]{000000} label}                              & {\color[HTML]{000000} 2975}                                                                          & \multirow{-2}{*}{{\color[HTML]{000000} 128 × 128}} & \multirow{-2}{*}{{\color[HTML]{000000} \begin{tabular}[c]{@{}l@{}}Cityscapes \\ training set ~\cite{55}\end{tabular}}}                              \\ \cline{2-6} 
{\color[HTML]{000000} }                                                                 & \multicolumn{1}{l|}{{\color[HTML]{000000} }}                                          & {\color[HTML]{000000} photo}                              & {\color[HTML]{000000} 400}                                                                           & {\color[HTML]{000000} }                            & {\color[HTML]{000000} }                                                                                                                          \\ \cline{3-4}
\multirow{-4}{*}{{\color[HTML]{000000} Urban Scenes}}                                   & \multicolumn{1}{l|}{\multirow{-2}{*}{{\color[HTML]{000000} facades photo $\leftrightarrow$ 
 label}}}     & label                                                     & 400                                                                                                  & \multirow{-2}{*}{{\color[HTML]{000000} 256 × 256}} & \multirow{-2}{*}{{\color[HTML]{000000} \begin{tabular}[c]{@{}l@{}}CMP Facade \\ Database ~\cite{56}.\end{tabular}}}                                \\ \hline
\end{tabular}
\end{table*}

\section{Experiments and Results}
Translation from MR–CT brain tumors  presented. First, pre-processing was performed on the raw MRI images. The transfer learning method was used to reduce the processing load and achieve successful results with a small amount of data.
\subsection{Dataset}
We used a real MR-CT dataset to train, develop and demonstrate the capabilities of our model. The dataset was obtained from ~\cite{5} and included 367 paired MR-CT brain images from 18 patients. To align the images, a mutual information rigid registration algorithm was used, and the MR images were corrected with the N3 bias field correction algorithm. The dataset contains both normal and abnormal(i.e., contains tumors) MR and CT images. With the RadiAnt DICOM viewer software, 2D image slices were taken from the MR-CT datasets and used in the experiments. The extracted images were then transformed into a PNG image data format with a resolution of 256 × 256 pixels. Estimation of the effectiveness and efficiency of the proposed model is done by training and testing datasets which are divided into we perform training on the 70\% of the given dataset and rest 30\% is used for the testing purpose. The model was tested using Google's Colab cloud service, TensorFlow 2.0, and the Python 3 framework. Our dataset is paired dataset which means; Each MR image in the dataset has a corresponding CT image, and vice versa. This pairing is essential for training and evaluating algorithms that aim to translate or convert MR images to CT images, or vice versa.

\subsection{Performance Metrics}
Full-Reference Image Quality Assessment (FR-IQA) is an automatic perceptual quality evaluation of a distorted image in comparison with a reference image. Images in FR-IQA are evaluated using PSNR and SSIM, two state-of-the-art evaluation metrics ~\cite{40}. These measures determine the level of distortion in synthetically created images. PSNR is the quickest and easiest method for gauging image quality. However, PSNR does not always correspond with what humans see and the quality of an image. It was suggested that more factors be used to overcome the limitation of PSNR metrics, specifically SSIM. UQI and VIF were also utilized in our analysis.

\begin{enumerate}
  \item \textbf{PSNR:} PSNR is widely utilized by many academics for image comparison and image synthesis because of its simplicity and ease of implementation, making it the most effective measuring tool to evaluate synthetic images ~\cite{41}. The PSNR evaluates how far off the generated image pixels are from the ground truth by measuring the amount of pixel loss. The paired images in the test dataset can only be evaluated using pixel loss-based measures like PSNR and SSIM. If the PSNR is high enough, a high-quality image will be produced. The equation for PSNR is given in \ref{e3}~\cite{40}:
    \begin{equation} \label{e3}
        PSNR=20.\log_{10}\left ( \frac{R^{2}}{MSE} \right) 
    \end{equation}

Where MSE is a Mean Square Error, and R is the maximum fluctuation in the input image data type. Based on \ref{e3}, if the input image has a double-precision floating-point data type, then R is 1. If it has an 8-bit unsigned integer data type, then R is 255 ~\cite{40}.

\item \textbf{SSIM:} Similarity between two images ($x$ and $y$) can be evaluated with the use of SSIM. To quantify or predict image quality using the SSIM index, an original uncompressed or distortion-free image must be used as a baseline. SSIM is a perception-based model that takes into account luminance masking and contrast masking as well as other crucial perceptual phenomena to account for how an image's quality is considered to have changed over time \ref{eq: SSMI}~\cite{42}.
\begin{equation}\label{eq: SSMI}
 SSIM(x,y)=\frac{{\left ( 2 \mu_{x}\mu_{y}+ C_{1} \right )+ \left (2 \sigma _{x_y}+C_{2}\right)}} 
{\left(\mu_{x^2}+\mu_{y^2}+{C_1})(\sigma_{x^2}+\sigma_{y^2}+{C_2}\right)}
\end{equation}

Where, $\mu_{x}$ and $\mu_{y}$ represent the mean of the particular image; $\sigma_{x^2}$ and $\sigma_{y^2}$ represent the standard deviation of the of the particular image; $\sigma _{x_y}$ represents the covariance between two images, and $C1$, $C2$ are constants set for avoiding instability.

\item \textbf{UQI:} In order to compare the distortion information between the original image and the distorted image, a mathematical metric called UQI is calculated, bypassing the need for a model of the human visual system. Loss of correlation, illuminance distortion, and contrast distortion all contribute to UQI. This metric is simple to compute and has several potential uses in the field of image processing ~\cite{40}. The formula for UQI is \ref{e4}~\cite{40}:
    \begin{equation} \label{e4}
        UQI=\frac{\colorbox{white}{$\sigma_{xy}$}}{\sigma_z \sigma_y}. \frac{2\widehat{x}\widehat{y}}{\widehat{x}^{2}+\widehat{y}^{2}}.\frac{2\sigma_x \sigma_y}{\sigma^{2}_x \sigma^{2}_y}
    \end{equation}
  The three components of the equation represent the loss of correlation, distortion of luminance, and distortion of contrast factors. Where $x$ is the original image and y is the generated image. $x$ and $y$ are defined in \ref{e5} and \ref{e6}~\cite{40}, respectively.
  \ref{e5}~\cite{40}:
    \begin{equation} \label{e5}
        \widehat{x}=\frac{1}{N}\sum i=1^{N}\left ( x_i \right )
    \end{equation}
    
    \begin{equation} \label{e6}
        \widehat{y}=\frac{1}{N}\sum i=1^{N}\left ( y_i \right )
    \end{equation}
  To perform multiplication and addition operations, $x$ and $y$ images must be square images with $N. N$.

\item \textbf{VIF:}
Natural Scene Statistics (NSS) and the idea of picture information retrieved by the human visual system form the basis of the Visual Information Feature (VIF) ~\cite{40}. In Figure ~\ref{fig_VIF components}, C represents the source image, D represents the distorted image, E represents the output of the Human Visual System (HVS) for the source image, and F represents the output of the HVS for the deformed image. Eq. \ref{e7}~\cite{40} represents VIF, where E and F represent the Reference and Distorted Images, respectively.

    \begin{equation} \label{e7}
        VIF = \frac{Distorted ~Image ~Information}{Reference~ Image ~Information} 
    \end{equation}

    \begin{figure}[!htbp]
    \centering
    \includegraphics[width=8cm, height=2cm]{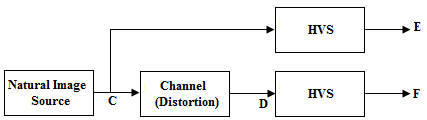}
    \caption{VIF components}
    \label{fig_VIF components}
    \end{figure}

\end{enumerate}

\subsection{Comparison of a pre-trained model}
\begin{figure*}[!htbp]
    \centering
    \includegraphics[width=18cm, height=20cm]{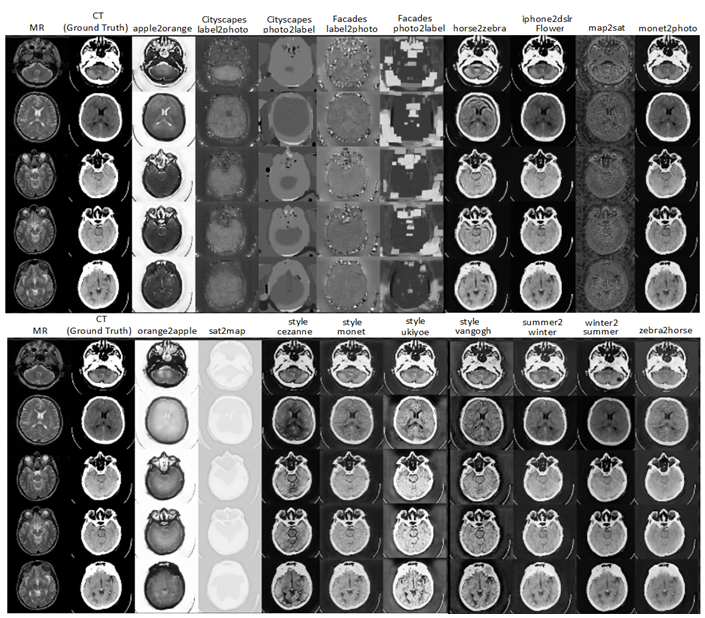}
    \caption{Output results of Magnetic Resonance to Computed Tomography (MR-CT) image translation from different generic pre-trained models.}
    \label{fig_MR_CT}
    \end{figure*}

\begin{figure*}[!htbp]
    \centering
    \includegraphics[width=18cm, height=20cm]{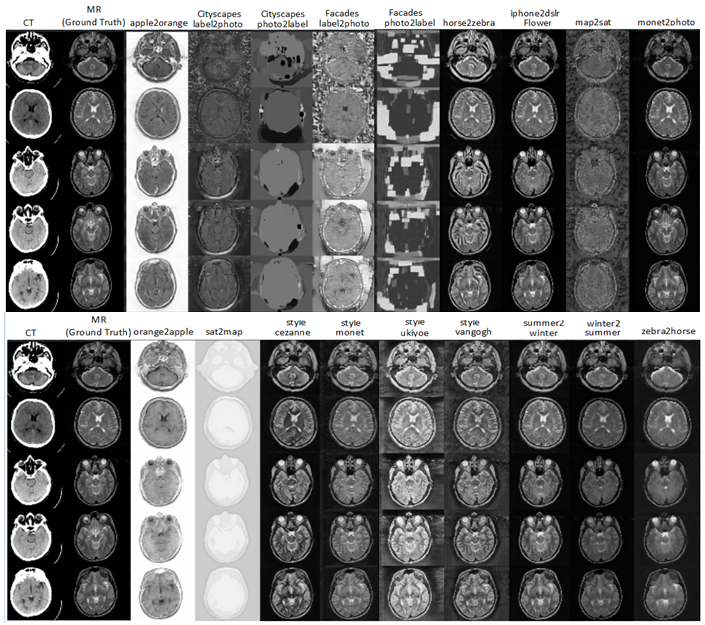}
    \caption{Output results of Computed Tomography to Magnetic Resonance (CT-MR) image translation from different generic pre-trained models.}
    \label{fig_CT_MR}
    \end{figure*}

In image synthesis tasks, the evaluation of generated images is crucial to assessing the quality of the model's output. The commonly used PSNR metric, which calculates the mean square error between the original and generated images, has limitations as it does not always correlate with human visual perception. Therefore, additional metrics such as SSIM, UQI, and VIF have been proposed to overcome these limitations. In this work, the performance of the model is evaluated using these metrics on both the paired and unpaired datasets. The experiments compared the bidirectional MR-CT synthesis using 18 pre-trained non-medical models, and the iphone2dslr\textunderscore flower model outperformed the others, resulting in the best score across most of the chosen metrics, with low standard deviation values indicating the stability of the results. Figures \ref{fig_MR_CT} and  \ref{fig_CT_MR} illustrate Bidirectional MR-CT Translation, using a pre-trained model. 

Table \ref{T2} and \ref{T3} present the outcomes of the PSNR, SSIM, UQI, and VIF metrics for the bidirectional MR-CT synthesis, covering both MR-to-CT and CT-to-MR translation using the entire pre-trained model.
However, the table shows the evaluation metrics of various pre-trained models on different image-to-image translation tasks. The metrics used are PSNR (Peak Signal-to-Noise Ratio), SSIM (Structural Similarity Index), UQI (Universal Quality Index), and VIF (Visual Information Fidelity). PSNR measures the quality of the reconstructed image compared to the original image, where a higher PSNR value indicates better image quality. SSIM measures the similarity between the reconstructed image and the original image based on structural information, where a higher SSIM value indicates better image quality. UQI measures the similarity between the reconstructed image and the original image based on human perception, where a higher UQI value indicates better image quality. VIF measures the information transfer between the reconstructed image and the original image, where a higher VIF value indicates better image quality. 

\begin{table*}[!htbp]
\caption{\label{T2} Image Quality Evaluation Metrics for MR-CT Translation}
\centering

\begin{tabular}{llllllllll}
\rowcolor[HTML]{FFFFFF} 
\multicolumn{1}{c}{\cellcolor[HTML]{FFFFFF}}                               & \multicolumn{9}{c}{\cellcolor[HTML]{FFFFFF}\textbf{MR-CT}}                                                                                                                                                                                                                                                                                                                                                                                                                                                                                                                       \\ \hline
\rowcolor[HTML]{FFFFFF} 
\multicolumn{1}{c}{\cellcolor[HTML]{FFFFFF}}                               & \multicolumn{1}{c}{\cellcolor[HTML]{FFFFFF}}                                             & \multicolumn{2}{c}{\cellcolor[HTML]{FFFFFF}\textbf{PSNR}}                                                           & \multicolumn{2}{c}{\cellcolor[HTML]{FFFFFF}\textbf{SSIM}}                                                           & \multicolumn{2}{c}{\cellcolor[HTML]{FFFFFF}\textbf{UQI}}                                                            & \multicolumn{2}{c}{\cellcolor[HTML]{FFFFFF}\textbf{VIF}}                                                            \\ \cline{3-10} 
\rowcolor[HTML]{FFFFFF} 
\multicolumn{1}{c}{\multirow{-2}{*}{\cellcolor[HTML]{FFFFFF}\textbf{NO.}}} & \multicolumn{1}{c}{\multirow{-2}{*}{\cellcolor[HTML]{FFFFFF}\textbf{Pre-trained Model}}} & \multicolumn{1}{c}{\cellcolor[HTML]{FFFFFF}\textbf{AVG}} & \multicolumn{1}{c}{\cellcolor[HTML]{FFFFFF}\textbf{STD}} & \multicolumn{1}{c}{\cellcolor[HTML]{FFFFFF}\textbf{AVG}} & \multicolumn{1}{c}{\cellcolor[HTML]{FFFFFF}\textbf{STD}} & \multicolumn{1}{c}{\cellcolor[HTML]{FFFFFF}\textbf{AVG}} & \multicolumn{1}{c}{\cellcolor[HTML]{FFFFFF}\textbf{STD}} & \multicolumn{1}{c}{\cellcolor[HTML]{FFFFFF}\textbf{AVG}} & \multicolumn{1}{c}{\cellcolor[HTML]{FFFFFF}\textbf{STD}} \\ \hline
\rowcolor[HTML]{FFFFFF} 
1                                                                          & apple2orange                                                                             & 28.63                                                    & $\pm$0.791                                               & 0.07                                                    & $\pm$0.052                                               & 0.02                                                     & $\pm$0.023                                               & 0.02                                                     & $\pm$0.034                                               \\
\rowcolor[HTML]{FFFFFF} 
2                                                                          & cityscapes\_photo2label,                                                                 & 27.67                                                    & $\pm$0.225                                               & 0.08                                                     & $\pm$0.033                                               & 0.01                                                     & $\pm$0.013                                               & 0.01                                                     & $\pm$0.024                                               \\
\rowcolor[HTML]{FFFFFF} 
3                                                                          & cityscapes\_label2photo                                                                  & 27.61                                                    & $\pm$0.275                                               & 0.14                                                     & $\pm$0.039                                               & 0.00                                                     & $\pm$0.003                                               & 0.00                                                     & $\pm$0.014                                               \\
\rowcolor[HTML]{FFFFFF} 
4                                                                          & facades\_label2photo                                                                     & 28.36                                                    & $\pm$0.389                                               & 0.07                                                     & $\pm$0.035                                               & 0.03                                                     & $\pm$0.033                                               & 0.03                                                     & $\pm$0.044                                               \\
\rowcolor[HTML]{FFFFFF} 
5                                                                          & facades\_photo2label                                                                     & 27.67                                                    & $\pm$0.769                                               & 0.07                                                     & $\pm$0.040                                               & 0.01                                                     & $\pm$0.013                                               & 0.01                                                     & $\pm$0.024                                               \\
\rowcolor[HTML]{FFFFFF} 
\cellcolor[HTML]{FFFFFF}6                                                  & \cellcolor[HTML]{FFFFFF}horse2zebra                                                      & 27.67                                                    & $\pm$0.503                                               & 0.36                                                     & $\pm$0.065                                               & 0.01                                                     & $\pm$0.011                                               & 0.01                                                     & $\pm$0.022                                               \\
\rowcolor[HTML]{FFFFFF} 
\cellcolor[HTML]{FFFFFF}7                                                  & \cellcolor[HTML]{FFFFFF}iphone2dslr\_flower                                              & 30.98                                                    & $\pm$0.119                                               & 0.65                                                     & $\pm$0.031                                               & 0.00                                                     & $\pm$0.000                                               & 0.01                                                     & $\pm$0.001                                               \\
\rowcolor[HTML]{FFFFFF} 
\cellcolor[HTML]{FFFFFF}8                                                  & \cellcolor[HTML]{FFFFFF}map2sat                                                          & 27.72                                                    & $\pm$0.142                                               & 0.16                                                     & $\pm$0.082                                               & 0.02                                                     & $\pm$0.021                                               & 0.02                                                     & $\pm$0.033                                               \\
\rowcolor[HTML]{FFFFFF} 
9                                                                          & monet2photo                                                                              & 29.54                                                    & $\pm$1.686                                               & 0.50                                                     & $\pm$0.076                                               & 0.03                                                     & $\pm$0.032                                               & 0.04                                                     & $\pm$0.043                                               \\
\rowcolor[HTML]{FFFFFF} 
10                                                                         & orange2apple                                                                             & 30.74                                                    & $\pm$1.054                                               & 0.05                                                    & $\pm$0.042                                               & 0.01                                                     & $\pm$0.012                                               & 0.01                                                     & $\pm$0.022                                               \\
\rowcolor[HTML]{FFFFFF} 
11                                                                         & sat2map                                                                                  & 26.82                                                    & $\pm$0.511                                               & 0.14                                                     & $\pm$0.031                                               & 0.00                                                     & $\pm$0.002                                               & 0.01                                                     & $\pm$0.014                                               \\
\rowcolor[HTML]{FFFFFF} 
12                                                                         & style\_cezanne                                                                           & 27.55                                                    & $\pm$0.207                                               & 0.38                                                     & $\pm$0.068                                               & 0.02                                                     & $\pm$0.022                                               & 0.03                                                     & $\pm$0.034                                               \\
\rowcolor[HTML]{FFFFFF} 
13                                                                         & style\_monet                                                                             & 27.56                                                    & $\pm$0.160                                               & 0.39                                                     & $\pm$0.068                                               & 0.02                                                     & $\pm$0.024                                               & 0.03                                                     & $\pm$0.036                                               \\
\rowcolor[HTML]{FFFFFF} 
14                                                                         & style\_ukiyoe                                                                            & 27.79                                                    & $\pm$0.288                                               & 0.33                                                     & $\pm$0.065                                               & 0.02                                                     & $\pm$0.024                                               & 0.03                                                     & $\pm$0.036                                               \\
\rowcolor[HTML]{FFFFFF} 
15                                                                         & style\_vangogh                                                                           & 27.80                                                    & $\pm$0.164                                               & 0.35                                                     & $\pm$0.069                                               & 0.03                                                     & $\pm$0.031                                               & 0.03                                                     & $\pm$0.042                                               \\
\rowcolor[HTML]{FFFFFF} 
16                                                                         & summer2winter\_yosemite                                                                  & 27.96                                                    & $\pm$0.461                                               & 0.30                                                     & $\pm$0.041                                               & 0.01                                                     & $\pm$0.012                                               & 0.01                                                     & $\pm$0.024                                               \\
\rowcolor[HTML]{FFFFFF} 
17                                                                         & winter2summer\_yosemite                                                                  & 29.22                                                    & $\pm$1.165                                               & 0.38                                                     & $\pm$0.084                                               & 0.04                                                     & $\pm$0.043                                               & 0.05                                                     & $\pm$0.054                                               \\
\rowcolor[HTML]{FFFFFF} 
18                                                                         & zebra2horse                                                                              & 27.81                                                    & $\pm$0.460                                               & 0.35                                                     & $\pm$0.060                                               & 0.02                                                     & $\pm$0.023                                               & 0.03                                                     & $\pm$0.034                                               \\ \hline
\end{tabular}
\end{table*}

\begin{table*}[!htbp]
\caption{\label{T3} Image Quality Evaluation Metrics for CT-MR Translation}
\centering

\begin{tabular}{
>{\columncolor[HTML]{FFFFFF}}l 
>{\columncolor[HTML]{FFFFFF}}l 
>{\columncolor[HTML]{FFFFFF}}l 
>{\columncolor[HTML]{FFFFFF}}l 
>{\columncolor[HTML]{FFFFFF}}l 
>{\columncolor[HTML]{FFFFFF}}l 
>{\columncolor[HTML]{FFFFFF}}l 
>{\columncolor[HTML]{FFFFFF}}l 
>{\columncolor[HTML]{FFFFFF}}l 
>{\columncolor[HTML]{FFFFFF}}l }
                                                       & \textbf{}                                                            & \multicolumn{8}{c}{\cellcolor[HTML]{FFFFFF}\textbf{CT-MR}}                                                                                                                                                                                                                                                                                                                                                                                                                            \\ \hline
\cellcolor[HTML]{FFFFFF}                               & \cellcolor[HTML]{FFFFFF}                                             & \multicolumn{2}{c}{\cellcolor[HTML]{FFFFFF}\textbf{PSNR}}                                                           & \multicolumn{2}{c}{\cellcolor[HTML]{FFFFFF}\textbf{SSIM}}                                                           & \multicolumn{2}{c}{\cellcolor[HTML]{FFFFFF}\textbf{UQI}}                                                            & \multicolumn{2}{c}{\cellcolor[HTML]{FFFFFF}\textbf{VIF}}                                                            \\ \cline{3-10} 
\multirow{-2}{*}{\cellcolor[HTML]{FFFFFF}\textbf{NO.}} & \multirow{-2}{*}{\cellcolor[HTML]{FFFFFF}\textbf{Pre-trained Model}} & \multicolumn{1}{c}{\cellcolor[HTML]{FFFFFF}\textbf{AVG}} & \multicolumn{1}{c}{\cellcolor[HTML]{FFFFFF}\textbf{STD}} & \multicolumn{1}{c}{\cellcolor[HTML]{FFFFFF}\textbf{AVG}} & \multicolumn{1}{c}{\cellcolor[HTML]{FFFFFF}\textbf{STD}} & \multicolumn{1}{c}{\cellcolor[HTML]{FFFFFF}\textbf{AVG}} & \multicolumn{1}{c}{\cellcolor[HTML]{FFFFFF}\textbf{STD}} & \multicolumn{1}{c}{\cellcolor[HTML]{FFFFFF}\textbf{AVG}} & \multicolumn{1}{c}{\cellcolor[HTML]{FFFFFF}\textbf{STD}} \\ \hline
1                                                      & apple2orange                                                         & 29.55                                                    & $\pm$0.781                                               & 0.16                                                    & $\pm$0.039                                               & 0.02                                                     & $\pm$0.034                                               & 0.12                                                     & $\pm$0.013                                               \\
2                                                      & cityscapes\_photo2label,                                             & 27.61                                                    & $\pm$0.105                                               & 0.06                                                    & $\pm$0.025                                               & 0.01                                                     & $\pm$0.024                                               & 0.11                                                     & $\pm$0.003                                               \\
3                                                      & cityscapes\_label2photo                                              & 27.27                                                    & $\pm$0.306                                               & 0.09                                                     & $\pm$0.027                                               & 0.00                                                     & $\pm$0.014                                               & 0.10                                                     & $\pm$0.007                                               \\
4                                                      & facades\_label2photo                                                 & 28.16                                                    & $\pm$0.272                                               & 0.09                                                    & $\pm$0.035                                               & 0.03                                                     & $\pm$0.044                                               & 0.13                                                     & $\pm$0.023                                               \\
5                                                      & facades\_photo2label                                                 & 27.62                                                    & $\pm$0.172                                               & 0.04                                                     & $\pm$0.029                                               & 0.01                                                     & $\pm$0.024                                               & 0.11                                                     & $\pm$0.003                                               \\
6                                                      & horse2zebra                                                          & 29.30                                                    & $\pm$0.662                                               & 0.40                                                     & $\pm$0.090                                               & 0.01                                                     & $\pm$0.022                                               & 0.11                                                     & $\pm$0.001                                               \\
7                                                      & iphone2dslr\_flower                                                  & 34.36                                                    & $\pm$0.072                                               & 0.83                                                     & $\pm$0.022                                               & 0.00                                                     & $\pm$0.001                                               & 0.10                                                     & $\pm$0.020                                               \\
8                                                      & map2sat                                                              & 27.77                                                    & $\pm$1.619                                               & 0.18                                                     & $\pm$0.036                                               & 0.02                                                     & $\pm$0.032                                               & 0.12                                                     & $\pm$0.011                                               \\
9                                                      & monet2photo                                                          & 33.49                                                    & $\pm$1.823                                               & 0.72                                                     & $\pm$0.069                                               & 0.03                                                     & $\pm$0.043                                               & 0.13                                                     & $\pm$0.022                                               \\
10                                                     & orange2apple                                                         & 31.02                                                    & $\pm$0.788                                               & 0.13                                                    & $\pm$0.034                                               & 0.01                                                     & $\pm$0.022                                               & 0.11                                                     & $\pm$0.001                                               \\
11                                                     & sat2map                                                              & 26.70                                                    & $\pm$0.778                                               & 0.08                                                     & $\pm$0.080                                               & 0.00                                                     & $\pm$0.013                                               & 0.10                                                     & $\pm$0.008                                               \\
12                                                     & style\_cezanne                                                       & 28.22                                                    & $\pm$0.329                                               & 0.43                                                     & $\pm$0.082                                               & 0.02                                                     & $\pm$0.033                                               & 0.12                                                     & $\pm$0.013                                               \\
13                                                     & style\_monet                                                         & 27.78                                                    & $\pm$0.269                                               & 0.43                                                     & $\pm$0.085                                               & 0.02                                                     & $\pm$0.035                                               & 0.13                                                     & $\pm$0.015                                               \\
14                                                     & style\_ukiyoe                                                        & 27.54                                                    & $\pm$0.097                                               & 0.29                                                     & $\pm$0.068                                               & 0.02                                                     & $\pm$0.035                                               & 0.13                                                     & $\pm$0.014                                               \\
15                                                     & style\_vangogh                                                       & 27.74                                                    & $\pm$0.160                                               & 0.36                                                     & $\pm$0.080                                               & 0.03                                                     & $\pm$0.042                                               & 0.13                                                     & $\pm$0.021                                               \\
16                                                     & summer2winter\_yosemite                                              & 30.16                                                    & $\pm$1.356                                               & 0.39                                                     & $\pm$0.061                                               & 0.01                                                     & $\pm$0.023                                               & 0.11                                                     & $\pm$0.002                                               \\
17                                                     & winter2summer\_yosemite                                              & 30.79                                                    & $\pm$0.715                                               & 0.44                                                     & $\pm$0.074                                               & 0.04                                                     & $\pm$0.054                                               & 0.14                                                     & $\pm$0.033                                               \\
18                                                     & zebra2horse                                                          & 27.88                                                    & $\pm$0.522                                               & 0.34                                                     & $\pm$0.078                                               & 0.02                                                     & $\pm$0.034                                               & 0.13                                                     & $\pm$0.013                                               \\ \hline
\end{tabular}
\end{table*}
The table shows that different pre-trained models perform differently on different image-to-image translation tasks. For example, the model "iPhone2dslr\textunderscore flower" has the highest PSNR and UQI values, indicating that it produces high-quality reconstructed images that are similar to the original images. On the other hand, the model "apple2orange" has the lowest PSNR and UQI values, indicating that it produces lower-quality reconstructed images that are less similar to the original images.
Based on the CT-MR table, it can be seen that the pre-trained model "iphone2dslr\textunderscore flower" has the highest PSNR value (34.355), indicating better image reconstruction. The model "summer2winter\textunderscore yosemite" has the highest SSIM (0.393) and UQI (0.393), indicating a high level of structural similarity and image quality between the two image domains. The model "zebra2horse" has the highest VIF (0.001113), indicating the highest level of visual information fidelity between the two image domains.

In the next section, we conducted a perceptual and validation study to further evaluate the performance of the pre-trained image-to-image translation models. We recognized that while metrics such as PSNR, SSIM, UQI, and VIF are helpful in quantitatively assessing the quality of the translated images, they do not necessarily capture the full extent of human perception and preferences. Therefore, we designed a study to include human evaluators who would provide qualitative feedback on the translated images, considering factors such as visual realism, image diversity, and overall aesthetic appeal. The validation study would help us to better understand the strengths and limitations of each pre-trained model and provide a more comprehensive evaluation of their performance.

\subsection{Perceptual study and validation}

The accuracy of the translated MR and CT pictures is assessed by showing them to two expert radiologists from JUH in a random order alongside the ground truth images. From a pool of 18 non-medical pre-trained models, the best one was selected for this analysis using PSNR, SSIM, UQI, and VIF metrics. With the primary goal of evaluating the images generated by the transfer learning model and comparing them to the ground truth images, the radiologists are shown a total of 240 MR images and 240 CT images with a resolution of 256 256 pixels. Radiologists are asked to determine which images are the ground truth and to assess the images' realism on a scale from 1 to 4, with 4 being the most realistic. 

Table \ref{T4} presents the results of the perceptual study evaluated by radiologists for MR-to-CT and CT-to-MR translations. The final column of this table shows the percentage of images classified as real by the radiologists over the total number of images.

\begin{table}[!htbp]
\caption{\label{T4} Results of Perceptual Study}
\centering
\begin{tabular}{cccclll}
\cline{1-4}
\multicolumn{4}{c}{\textbf{MR-CT}}                                                                &  &  &  \\ \cline{1-4}
\textbf{Model}               & \textbf{Mean}        & \textbf{Std}         & \textbf{Real\%}      &  &  &  \\
\textbf{iphone2dslr\_flower} & 3.79                 & 0.21                 & 97.91\%                &  &  &  \\
\textbf{Ground   truth}      & 3.85                 & 0.23                 & 98.58\%                &  &  &  \\ \cline{1-4}
\multicolumn{4}{c}{\textbf{CT-MR}}                                                                &  &  &  \\ \cline{1-4}
\textbf{Model}               & \textbf{Mean}        & \textbf{Std}         & \textbf{Real\%}      &  &  &  \\
\textbf{iphone2dslr\_flower} & 3.69                 & 0.21                 & 97.7\%                 &  &  &  \\
\textbf{Ground   truth}      & 3.54                 & 0.23                 & 98.37\%                &  &  &  \\
\multicolumn{1}{l}{}         & \multicolumn{1}{l}{} & \multicolumn{1}{l}{} & \multicolumn{1}{l}{} &  &  &  \\
\multicolumn{1}{l}{}         & \multicolumn{1}{l}{} & \multicolumn{1}{l}{} & \multicolumn{1}{l}{} &  &  &  \\
\multicolumn{1}{l}{}         & \multicolumn{1}{l}{} & \multicolumn{1}{l}{} & \multicolumn{1}{l}{} &  &  &  \\
\multicolumn{1}{l}{}         & \multicolumn{1}{l}{} & \multicolumn{1}{l}{} & \multicolumn{1}{l}{} &  &  & 
\end{tabular}
\end{table}

The given table shows the performance of the iphone2dslr\textunderscore flower pre-trained model in MR-to-CT translations. It is mentioned that the model achieved a mean score of 3.79, which reflects its good performance in generating CT images from MR images. Also, the table goes on to state that this pre-trained model also performs well in CT-to-MR translations, achieving a mean score of 3.69.

Moreover, the success rate for CT images is 97.91\%, while for MR images, it is 97.7\%, which indicates that the radiologist was convinced that they were real scans.

We compute Lin’s concordance correlation coefficients (CCC) for the results of the perceptual study, which quantifies the agreement between the ground truth images and the generated images.

Table \ref{T5} presents Pearson’s correlation coefficient ($\rho C$) and accuracy ($C\beta$) for both MR-to- CT and CT-to-MR translation. The results indicate a high or medium correlation between ground truth and the generated images for all radiologists.

\begin{table}[!htbp]
\centering
\caption{\label{T5} Agreement of Generated and Ground-Truth Images}
\begin{tabular}{lllllll}
\cline{1-5}
\multirow{2}{*}{pre-trained   model} & \multicolumn{2}{l}{MR-CT} & \multicolumn{2}{l}{CT-MR} &  &  \\
                                     & ($\rho C$)        & ($C\beta$)        & ($\rho C$)        & ($C\beta$)        &  &  \\ \cline{1-5}
Radiologist1                         & 0.69        & 0.99        & 0.88        & 0.94        &  &  \\
Radiologist2                         & 0.97        & 1.00        & 0.60        & 0.97        &  &  \\
\end{tabular}
\end{table}

\subsection{Latent Space}
\begin{figure*}[!htbp]
\centering
\includegraphics[width=13 cm, height=10 cm]{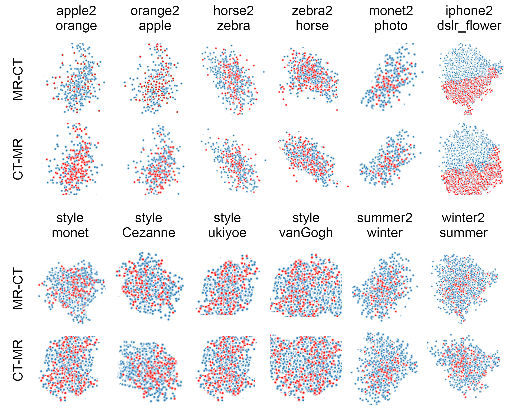}
\caption{Latent space visualization using different pre-trained generic models for MR-CT and CT-MR image translation.}
\label{fig_LatentSpace}
\end {figure*}

Latent space, in the context of machine learning and data representation, refers to a lower-dimensional, abstract space where essential information and patterns of data are encoded~\cite{48}. It is a transformative concept that allows complex data, such as images or text, to be condensed into a more manageable and informative form. In the case of medical image translation from magnetic resonance imaging (MRI) to computed tomography (CT), the concept of latent space becomes particularly valuable~\cite{46}. This process effectively bridges the gap between the two imaging techniques, capturing common features and reducing the impact of modality-specific differences~\cite{48}. Latent space, often created through dimensionality reduction techniques like Principal Component Analysis (PCA) or t-distributed stochastic neighbor embedding (t-SNE), allows us to represent complex and high-dimensional data in a more comprehensible manner. In various fields, from machine learning to biology to social sciences, latent space plays a pivotal role in revealing hidden patterns, relationships, and clusters within data~\cite{47}.
In our work, a latent space was created for the pre trained models that performed well, which is :apple2orange, orange2apple, horse2zebra, zebra2horse, style\textunderscore monet, style\textunderscore Cezanne, style\textunderscore ukiyoe, style\textunderscore vanGogh, summer2winter \textunderscore yosemite, winter2summer\textunderscore yosemite, monet2photo and iphone2dslr\textunderscore flower). Figure~\ref{fig_LatentSpace} provides a visual representation of the latent space, offering a compelling illustration of how the iphone2dslr\textunderscore flower translation task excelled in correctly separating the distinctive characteristics of the input images. 
The latent space in figure ~\ref{fig_LatentSpace} has two colors, these colors represent distinct features or characteristics extracted from the input images. The simplification into two colors implies a high-level abstraction or compression of information and the base model was trained for two classes, where red represents CT images and blue represents MR images.
However, upon closer inspection, it is clear that colors in most of pre-trained models overlap within the existing space, indicating the models' difficulty in effectively separating or distinguishing input image elements during training. This difficulty may stem from various factors such as variability within the data, complex relationships between MRI and CT features, or limitations within the model structure.

Conversely, in the case of the iphone2dslr\textunderscore flower model, clear color separation was observed within the latent space. This indicates that the model has successfully learned to distinguish between distinct features or patterns found in MR and CT images, indicating effective training. This analysis indicates that the model has encoded essential information within the latent space in a way that preserves the fundamental differences between MR and CT methods. Thus, when generating cross-sectional images through inverse MR imports, the model translation and underlying features can be better understood and interpreted.

Clearer separation within the latent space implies more robust and accurate performance, as the model can more effectively differentiate between the diverse effects of each imaging modality. This underscores the importance of choosing a model that relies not only on visual observations, but on an in-depth analysis of the data it produces.

\section{Discussion:}
\begin{figure}[!htbp]
    \centering
    \includegraphics[width=8cm, height=5cm]{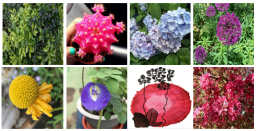}
    \caption{Sample images from generic iphone2dslr\textunderscore flower pre-trained model.}
    \label{iphone}
    \end{figure}

\begin{figure}[!htbp]
    \centering
    \includegraphics[width=8cm, height=5cm]{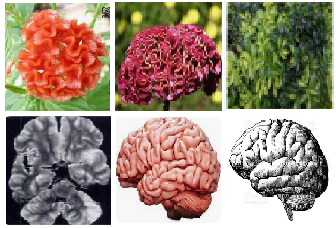}
    \caption{Physiological characteristic similarities between generic iphone2dslr\textunderscore flower images and brain images.}
    \label{iphone_brain}
    \end{figure}

In Fig. 5, a comprehensive analysis of the outcomes reveals a consistent correlation with the findings presented in Fig. 3, which showcases the diverse results obtained from the pre-trained model. Notably, a subset of these models, including facades photo2label and facades label2photo, exhibited erroneous outcomes attributed to their primary specialization in region identification and labeling. Moreover, the sat2map variant also demonstrated significantly divergent results from the desired outcomes. As a result of our meticulous examination, we arrive at the conclusion that the crucial determinant for achieving satisfactory results lies in the judicious selection of a pre-trained model that is specifically tailored to a dataset possessing properties closely resembling the new images used for the model's training. This underscores the pivotal role played by appropriate model selection in ensuring the successful training of the model.
The iphone2dslr\textunderscore flower model stands out as the premier pre-trained non-medical model for image translation, and it has also exhibited exceptional capabilities in the medical field, particularly in translating MRI images into CT scans, achieving top scores in all four evaluation scales. This pre-trained generative model was trained on a vast image dataset, the iphone2dslr\textunderscore flower dataset, which proves to be an ideal choice for transfer learning in photomontage tasks. This dataset comprises images of flowers captured by both iPhone and DSLR cameras, allowing the training of models capable of generating high-resolution flower images that closely resemble the quality of those taken with a DSLR camera. The model's training involved a diverse range of images, including flowers that share visual similarities with certain structures in the human body, such as blood vessels \cite{51}. Additionally, some flower types exhibit compact corals with intricate convoluted surfaces reminiscent of brain structures, making this dataset especially valuable for models aiming to reproduce the intricate details and textures found in these floral specimens. The petals displayed a striking resemblance to the inner tissues of the brain, exhibiting a repeating pattern that mirrors the distinctive torsions and folds present in cerebral structures~\cite{32}. This intriguing similarity between the model's results and brain anatomy highlights the potential of this approach to capture complex biological patterns and structures. The model's ability to capture such fine details indicates a close match between the training data and brain features, ultimately contributing to its exceptional performance in the realm of brain image analysis. Figure \ref{iphone} showcases some of the flowers utilized in the iphone2dslr\textunderscore flower dataset, further illustrating the dataset's diversity and relevance for the model's training. and Figure \ref{iphone_brain} Resemblance between Brain Images and iPhone Data. The images showcase striking similarities between brain structures and the data generated by the iPhone model. The intricate patterns and textures found in certain types of flowers captured by the iPhone camera demonstrate a captivating resemblance to brain tissues \cite{52}, highlighting the potential of the model in capturing complex biological structures.

The model can learn the features and patterns present in the images, which can then be used to generate new images. The process includes loading the pre-trained model, preparing the input image, using the pre-trained model to generate the faked image, post-processing, and evaluating the generated image using metrics such as MSE and SSIM ~\cite{38}. It is worth mentioning that the iphone2dslr\textunderscore flower model is a conditional generative model that can generate realistic flower images from a given input image, it is not directly designed for medical image translation, but with some adjustments and fine-tuning it could be used to generate images that are similar in appearance and information content to the original images.

\section{Conclusion:}
It has been demonstrated by our study that the potential of pre-trained models for the task of medical image translation, specifically for MRI-CT and vice versa, can be seen. Eighteen pre-trained models were used from a range of publicly available models, including models trained on natural images and models specifically designed for medical image translation tasks. It was found that these models performed well on all four commonly used image quality metrics: PSNR, SSIM, UQI, and VIF. It was also suggested that models that had been specifically designed for medical image translation tasks and trained on a large dataset of medical images performed better than models that had been trained on general image translation tasks.
However, it is essential to note that to obtain the best results for specific medical image translation tasks, such as brain, lung, or cardiac images, it may be necessary to fine-tune the models on a dataset of medical images specific to the task at hand and use a larger dataset. This way, the efficiency and accuracy of diagnosis and treatment in clinical practice can be improved. The best performance across all metrics was achieved by the 'iphone2dslr\textunderscore flower' model, thus making it the most suitable model for this task. The potential of transfer learning in medical imaging and the effectiveness of the 'iphone2dslr\_ flower' model have been demonstrated by our findings.
In general, the promising capabilities of pre-trained models for medical image translation tasks have been confirmed by this study, and it has been suggested that these techniques could be implemented in the near future for clinical use.

\section*{Acknowledgment}
The authors would like to thank Dr. Isra’a Almallahi and Dr. Ruba Braik for their assistance with the perceptual study. This work was supported by The Scientific Research and Innovation Support Fund, Ministry of Higher Education and Scientific Research (Amman, Jordan) [grant number ICT/1/03/2021]. 

\printbibliography
\end{document}